\documentclass[letterpaper]{article}
\usepackage{natbib,alifeconf}  
\usepackage{bm}

\usepackage[hidelinks]{hyperref}
\usepackage{comment}
\usepackage{color}

\usepackage{optidef}
\usepackage{ulem}
\DeclareMathOperator*{\argmax}{argmax} 
\usepackage{amssymb}
\usepackage{stackengine}

\usepackage[caption=false,font=normalsize]{subfig}

\usepackage{todonotes}
\usepackage[shortlabels]{enumitem}

\definecolor{cambridgeblue}{rgb}{0.64, 0.76, 0.68}

\usepackage{etoolbox}

\newtoggle{ShowNames}
\toggletrue{ShowNames}

\title{Can We Tell if ChatGPT is a Parasite? \\Studying Human--AI Symbiosis with Game Theory}

\iftoggle{ShowNames}{
    \author{Jiejun Hu-Bolz \and James Stovold \\\mbox{} Lancaster University Leipzig, Nikolaistra\ss{}e 10, 04109 Leipzig, Germany\\\texttt{\{j.hu14, j.stovold\}@lancaster.ac.uk}}
}
{%
    \author{Author 1$^1$ \and Author 2$^1$ \\ \mbox{} $^1$Address Line 1 \\ \texttt{email@college.edu}}
}

\begin{document}
\maketitle

\begin{abstract}

This work asks whether a human interacting with a generative AI system can merge into a single individual through iterative, information-driven interactions. We model the interactions between a human, a generative AI system, and the human's wider environment as a three-player stochastic game. We use information-theoretic measures (entropy, mutual information, and transfer entropy) to show that our modelled human and generative AI are able to form an aggregate individual in the sense of \citet{krakauer2020information}. 

The model we present is able to answer interesting questions around the symbiotic nature of humans and AI systems, including whether LLM-driven chatbots are acting as parasites, feeding on the information provided by humans.

\end{abstract}

\noindent This paper is accepted in ALife 2025 \\
\noindent Data/Code available at: 
\iftoggle{ShowNames}{\url{https://github.com/JunHu-Bolz/ALife_sgame.git}}{\url{https://github.com/JunHu-Bolz/ALife_sgame.git}}

\section{Introduction}

With the widespread adoption of Large Language Models (LLMs), humans are interacting with 
AI systems more easily and more often than ever before. The text-based interface makes interaction with 
AI systems simpler and reduced the barrier to using them for simple tasks. Chat-based GenAI systems (e.g.\ ChatGPT and DeepSeek) are regularly used to get quick answers to everyday questions, assist learning, brainstorm new ideas, and then apply these ideas to the real world. 

This increased interaction between humans and AI raises interesting questions about the nature of this interaction, the role of AI in human creativity, the dynamics of how AI systems are adapting to better answer human questions, and how the human is impacted by these changes in its wider behaviour. In this paper, we use LLM-based generative AI as a concrete, widely-familiar example to shape the discussion and illustrate key dynamics.

The field of distributed cognition~\citep{hollan_distributedcognitiontoward} has long pondered the question of how cognitive processes are impacted by working in close proximity with others, and how the tools and environment surrounding humans has an impact on the way they think and the choices they make. Modern AI systems are different in the way we typically interact with them (i.e.\ through a text-based interface, in an iterative manner) and are different in that they are able to quickly adapt to the user. 

There are some fundamental issues with this type of iterative interaction between a user, their wider environment, and an AI system that are worthy of consideration:
\begin{itemize}[noitemsep]
    \vspace{-2mm}
    \item LLMs are statistical guessers, not truth-tellers.
    \item Users and environments are dynamic and noisy in decision making.
    \item Errors cascade due to interdependent feedback and iterative interaction.
    \vspace{-2mm}
\end{itemize}
Understanding the impact of these points on the dynamics of human--AI--environment interaction is increasingly important as AI systems become more integrated into daily life. 

Human--AI interaction is a well-studied sub-field of Human--Computer Interaction (HCI). With the advent of generative AI systems, human--AI interaction has followed suit, studying the ability of these systems in aiding designers~\citep{swearngin_scoutrapidexploration} and bolstering creativity~\citep{walton_evaluatingmixedinitiative}. The studies of \citet{lee_guicompguidesign} and \citet{duan_generatingautomaticfeedback}, however, suggest that AI systems tend to be useful early on but become less useful over time. The iterative interaction between human and a non-adaptive AI system could be susceptible to this. In our work, we consider an \textit{adaptive} AI system which learns from its interactions with the human, modelling this as a multi-agent system to study a variety of possible scenarios, and paving the way to modelling societal-scale interactions.

Different frameworks have been developed to explore mutual understanding and co-adaptation between humans and AI~\citep{huhuman}. To understand the relationship between prompt and response in human--AI interaction, \citet{zhang2024understanding} propose a prompt-response concept (PRC) model that explains how LLMs generate responses and helps understand the relationship between prompts and response uncertainty---the uncertainty of responses decreases as the prompt's informativeness increases, similar to epistemic uncertainty. \citet{bhargava2023s} formalize LLMs as a class of discrete stochastic dynamical systems and explores prompt engineering through the lens of control theory. The empirical results show that short prompt sequences (up to 10 tokens) can significantly alter the likelihood of specific outputs, even making the least-likely tokens become the most-likely ones. This indicates that information density might serve a key role in prompt engineering.

Information-theoretic frameworks are adopted to analyse interaction between humans and AI systems. \citet{abbasi2024believe} present an information-theoretic approach for quantifying the epistemic uncertainty in LLM outputs, which can be useful for improving the reliability and trustworthiness of these models. \citet{wang2022mutual} propose the Mutual Theory of Mind (MToM) which posits that effective human--AI communication arises when both parties develop models of each other's intentions and behaviours. This work helps to guide the design of AI systems that are capable of iterative and adaptive communication with humans. \citet{xie2024measuring} introduce a framework that quantifies the human contribution by calculating the mutual information between human inputs and AI-generated outputs. This method provides a nuanced understanding of the collaborative dynamics in content creation, distinguishing varying degrees of human involvement. \citet{krakauer2020information} provide an information-theoretic model of individuality, demonstrating how the full spectrum of individuality can be measured based on interactions between a system (cells, animals, societies) and their environment. 
 
Unlike prior studies, this work aims to capture user adaptation, to quantify the role of the human's environment, and uses information-theoretic metrics to elucidate the impact of choices during AI system design---offering new insights into co-adaptation, predictability, and optimal conditions for effective ongoing human--AI collaboration. The main contributions of this work include:
\begin{itemize}[noitemsep]
    \vspace{-2mm}
    \item This work models the interaction between a human, AI and the human's environment, using a user and GenAI as an example. This holistic approach captures the complex interdependencies and feedback loops that occur in real-world scenarios, providing deeper insights into how each component influences and adapts to the others over time.
    \item The proposed temporal model reveals how mutual information and system uncertainty develop, offering a more nuanced understanding of the co-adaptive processes between users and GenAI systems.
    \item The proposed model demonstrates that a user--GenAI collaboration can result in both benefits (e.g. knowledge expansion) and risks (e.g.\ parasitic relationship). 
        \vspace{-2mm}
\end{itemize}

\section{Scenario}

Our intent is to model the interaction between humans and AI systems in a rigorous way, using information-theoretic measures to answer fundamental questions about this interaction. In this paper, we present the first step towards this, with a single-human, single-machine (equipped with AI), single-environment model. We define the environment as the physical space in which the user is trying to complete a task, along with any physical context directly related to that task. Understanding the task itself is not enough---we also need to capture the environment as the user perceives and reports it. Some tasks (using machine equipped with LLM as an example), such as proofreading, fact-checking, summarising content, or drafting emails, require little or no environmental information. In contrast, tasks involving in-situ elements and ongoing human–AI interaction---such as driving assistance---depend heavily on the user's input about the current environment. Even if an LLM has been trained on environmental information, it cannot perceive real-time changes; the user's timely, accurate input is essential for generating appropriate in-situ responses.

\begin{figure}[t]\vspace{-2mm}
    \centering
    \includegraphics[width=0.85\linewidth]{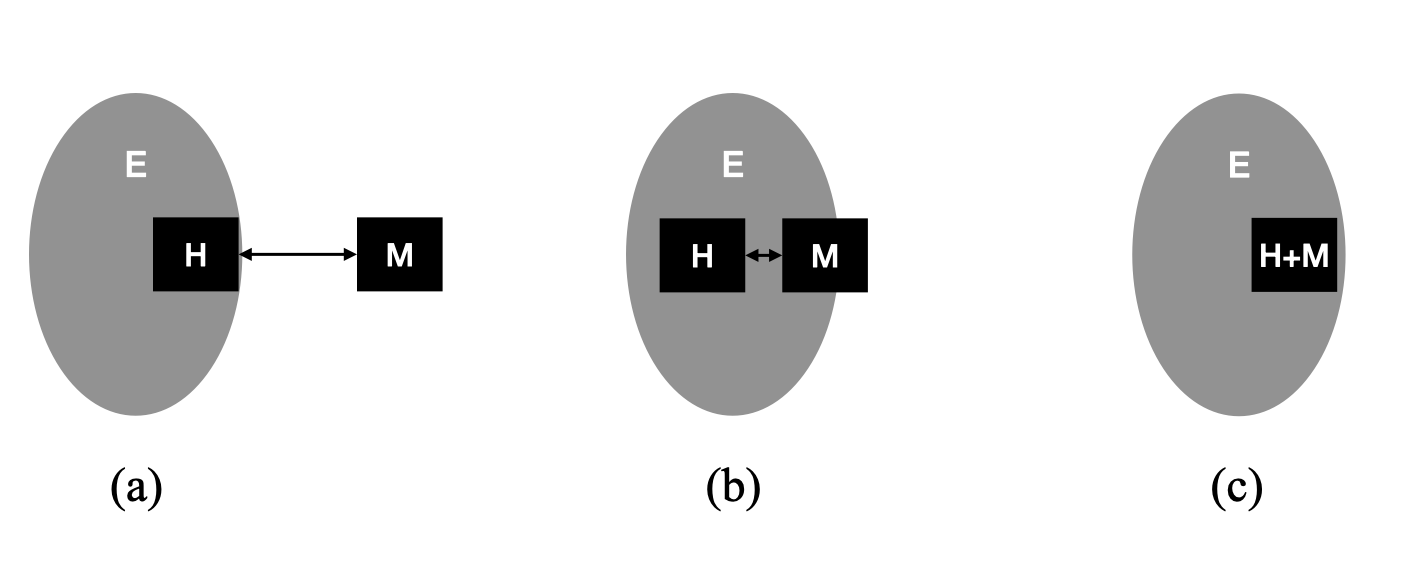}\vspace{-5mm}
    \caption{When human and machine become one individual }        
    \label{fig:bean}
\end{figure}

The scenario under consideration is depicted in Fig.~\ref{fig:bean}. It shows three stages of a single human interacting with a single machine. The human is embedded into an environment, and the machine is separate from the human's environment. The human can extract information from the environment, but this comes at a cost (pertaining to the cognitive effort involved), or the human can query the machine (which is cheap, but the likelihood of getting a high-quality answer is lower). 

Consider the scenario in Fig.\ \ref{fig:bean}(a). The human is embedded into its environment and has a query for the machine. As the human interacts with the machine, both the human and machine will collect knowledge on each other. With each prompt from the human to the machine, information about the environment will flow through the human to the machine. The machine will adapt based on this new knowledge.

As the human uses the machine more, the machine learns more about the human and about the human's view of the environment. In this period, the machine and human are more tightly coupled (Fig.\ \ref{fig:bean}(b)); the human is increasingly reliant on the machine rather than the environment for answers/information. The machine is still only guessing, but the human needs to expend more effort to get an answer from the environment than from the machine. 

This is a vicious cycle: as the human gets less efficient at working with the environment, the cognitive cost of working with the environment increases, so the human goes to the machine more often, and so on. The machine is learning about the human and about the human's perceptual worldview (its perspective of the environment), and can tailor its responses accordingly. Each round of prompting and response couples the human's cognition (e.g.\ goals, biases) with the machine's statistical priors (e.g.\ training data, inference rules). This makes the machine's guesses increasingly plausible to the human, and the only mechanism available for the human to check the answers is by going to the environment. 

By Fig.\ \ref{fig:bean}(c), the human and machine are inseparable---they have become one individual. The machine is still only getting information about the environment through the human, but the human is relying so heavily on the machine that they regularly use the machine for 
daily tasks, 
allowing the machine to continually obtain environmental and contextual information.
The increased cognitive load of using the environment is such that the human is now adapting their prompts to get better responses from the machine which, in turn, tells the machine more about the human and the environment. 

At this stage, the human and machine are in a symbiotic relationship, with the machine benefitting (informationally) and the human losing the ability to work effectively with the environment---characteristic of a parasitic relationship. 

We model this scenario as a stochastic game. Using information theory, we analyse the conditions under which the human and machine can be considered a single individual---specifically, by evaluating whether they form an aggregate system that propagates information from past states into the future~\citep{krakauer2020information}. With our proposed model, we aim to assess the emergence of symbiotic relationships between human and machine based on information flow and collective dynamics.
\vspace{-2mm}
\section{Model of Human-AI Interaction}
\begin{figure}[h]\vspace{-2mm}
    \centering
    \includegraphics[width=0.8\linewidth]{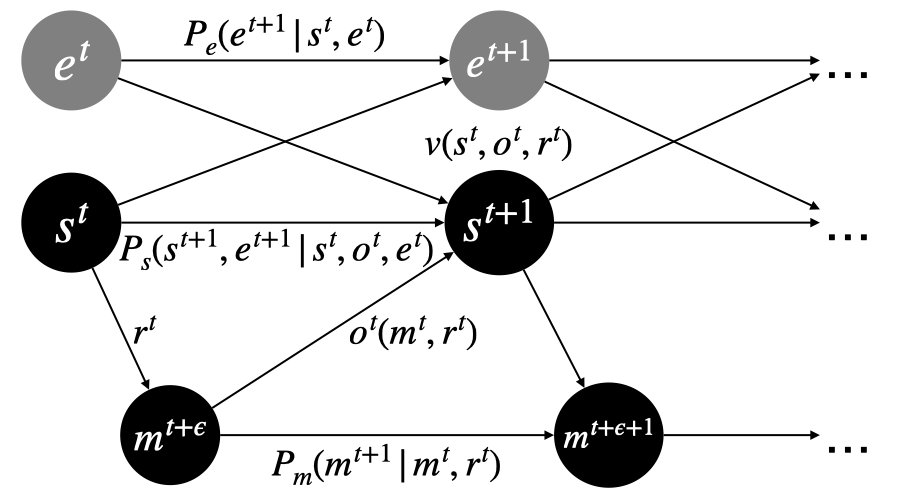}\vspace{-4mm}
    \caption{Model of user-AI-environment}
    \label{fig: model}
\end{figure}

In the proposed model, we illustrate human-AI interaction through the example of a user interacting with GenAI. In the remainder of the paper, we interchangeably use ``human'' to refer to the user and ``GenAI'' to refer to machine. We consider a single user interacting with a machine equipped with GenAI to answer a closed-ended question through multiple rounds of interaction. We define the current knowledge as the user state, $s^t \in \mathcal{S}$ and use this knowledge to generate the prompt, $r^t \in \mathcal{R}$ at time $t$. In this paper, we assume that the quality of the prompt is positively related to the user's knowledge. There is communication latency $\epsilon$ between the user and the machine (it is assumed that $\epsilon$ is small enough as to be considered negligible in the analysis). After submitting the prompt to the machine, the machine uses its foundation model with state $m^t$ and the prompt $r^t$ to generate a reply. We define the reply as the observation $o^t(m^t, r^t)$, as shown in Fig.~\ref{fig: model}.
After receiving the reply, the user aims to maximise the information between their current knowledge $s^t$ and observation $o^t$ to answer the closed-ended question. 

The user also considers environmental influences, such as social and physical environments, which can affect decision-making. The environmental state is denoted as $e^t\in \mathcal{E}$. We can define the transition of the user's current knowledge as 
\begin{equation}
    s^{t+1} = f (s^t, o^t) + e^t
    \label{eq: trans_user}
\end{equation}
where function $f(\cdot)$ is an evaluation of the current state and observation. We then define the transition probability of the next user and environment states as 
\begin{equation}
    P(s^{t+1}, e^{t+1}|s^t, o^t, e^t) = P_s(s^{t+1}|s^t, o^t, e^t)P_e(e^{t+1}|s^t, e^t)
\end{equation}
where $P_s$ and $P_e$ are transition probability of user state $s$ and environment state $e$, respectively. Note that the user future state is affected by the current state, response of the machine, and the current environment state, and the environment future state relies on the current environment state and user's state.
We further define the immediate payoff of the user as 
\begin{equation}
    v^t = \alpha u(s^t, r^t) - o^t
    \label{eq: reward_user}
\end{equation}
where $\alpha$ is a positive weight, $r^t$ is the action of the user, which is a function of $s^t$, $u$ is a concave function that represents user's satisfaction (reflecting the principle of diminishing marginal returns), and $o^t$ is the energy cost of observing the reply from the machine. We can further define the state transition probability of the machine as $P_m(m^{t+1}| m^t, r^t)$.

In this interaction of the user and the machine, the policy $\pi$ of the user is a plan to play the game. Here, $\pi = (\pi^1, \pi^2, ..., \pi^t, ...)$, where $\pi^t$ is the decision mapping the history of the interaction up to time $t$ to the action of selecting the value function: $\pi^t: h^t \in \mathcal{R}$. we define $h^t = (s^1, r^1, o^1, ...)$
Hence, we define the long-term discounted average payoff of the user as 
\begin{align}
    V(s, \pi)&= (1-\beta)\sum_{t=1}^T \beta^{t-1} v^t \nonumber \\
    &= (1-\beta)\sum_{t=1}^T \beta^{t-1} [\alpha u^t(s^t, r^t) + o^t]
\end{align}
where $\beta$ is a discount factor. The best response of the user is further defined as follows
\begin{align}
    r(s, \pi^*) &= \argmax_{r \in \mathcal{R}} (1-\beta) [\alpha u^t(s^t, r^t) + o^t] \nonumber \\
    &+ \beta [\sum_{s', e'} P(s', e'|s, o, e) P(m') V(s', \pi^*)]
    \label{eq: user_opt}
\end{align}
where $s'$, $m'$, and $e'$ are future states and $\pi^*$ is the optimal strategy. We cast the user--GenAI--environment loop as a stochastic game---where states capture user context, actions are based on knowledge, transitions follow observed feedback, and the optimal policy is the action-selection rule that maximises expected cumulative reward as given by the Bellman optimality equation.   
\begin{figure*}[h!]
\centering
\subfloat[\footnotesize Unbiased knowledge ($\alpha=1$)]{\includegraphics[width=0.31\linewidth]{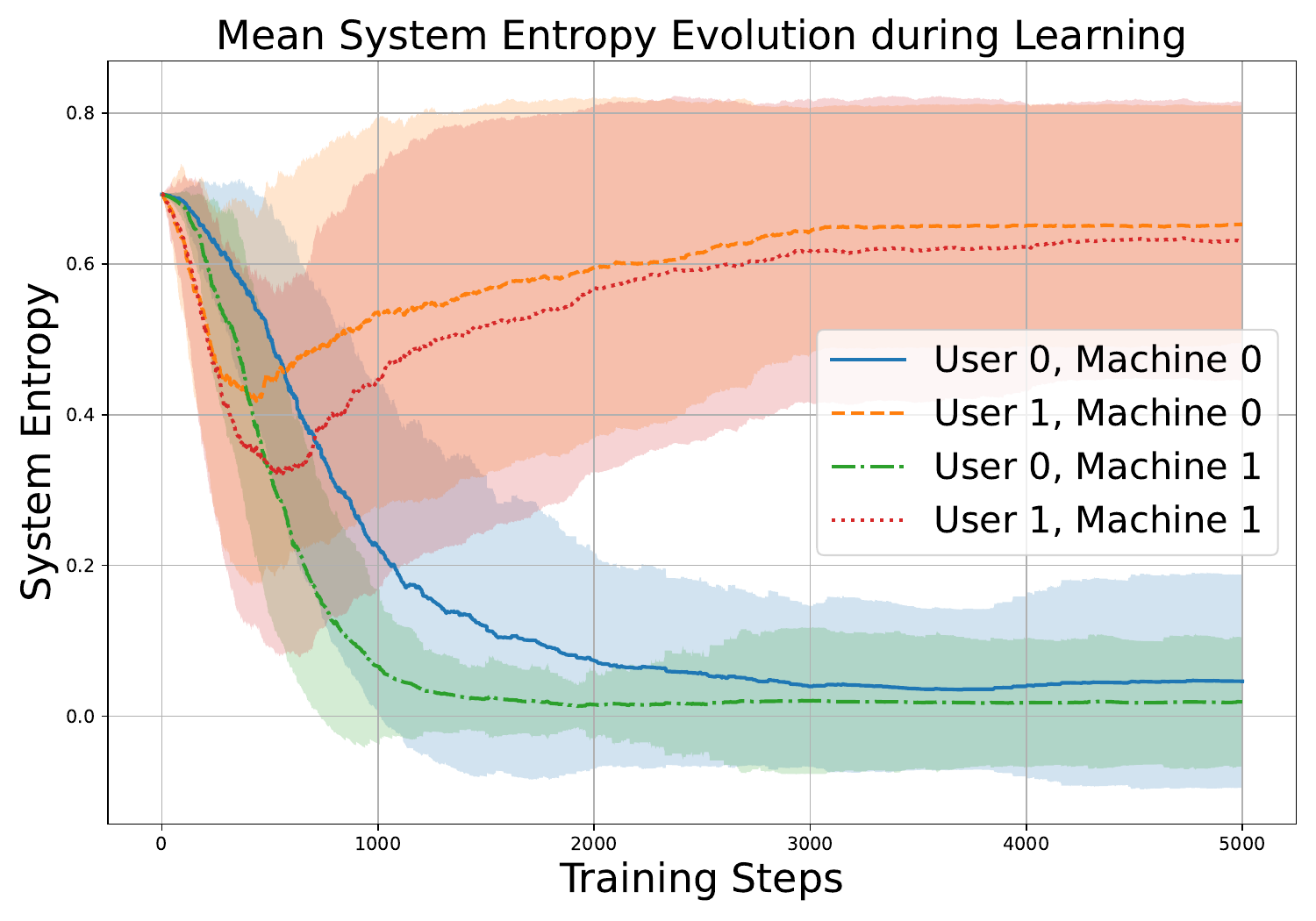}%
\label{fig: u_s_h}}
\hfil
\subfloat[\footnotesize Biased knowledge ($\alpha=1$)]{
\label{fig: u_s_h_skewed}
\includegraphics[width=0.31\linewidth]{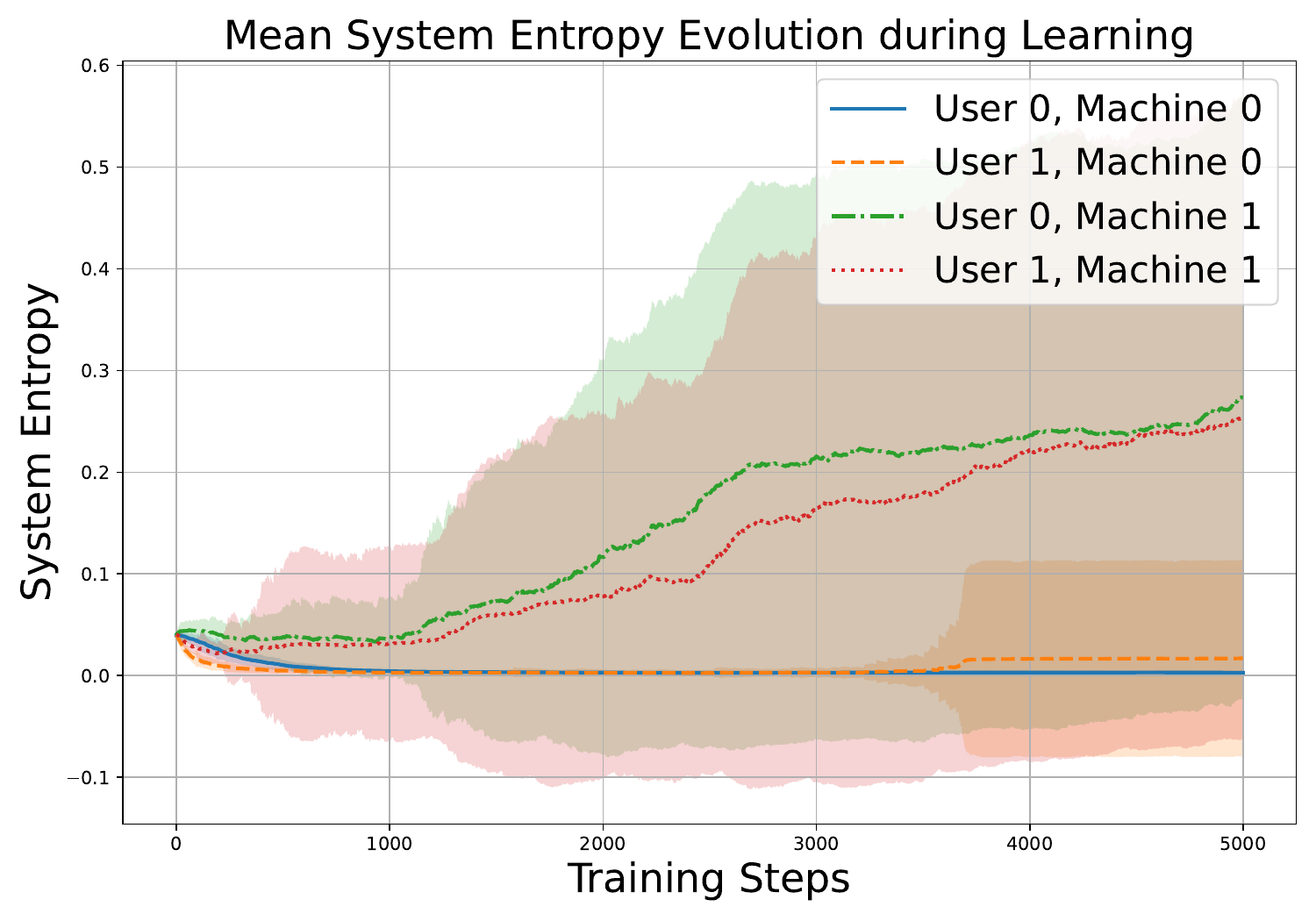}%
}
\hfil
\subfloat[\footnotesize With higher reward $\alpha=2$]{
\label{fig: u_s_h_reward}
\includegraphics[width=0.31\linewidth]{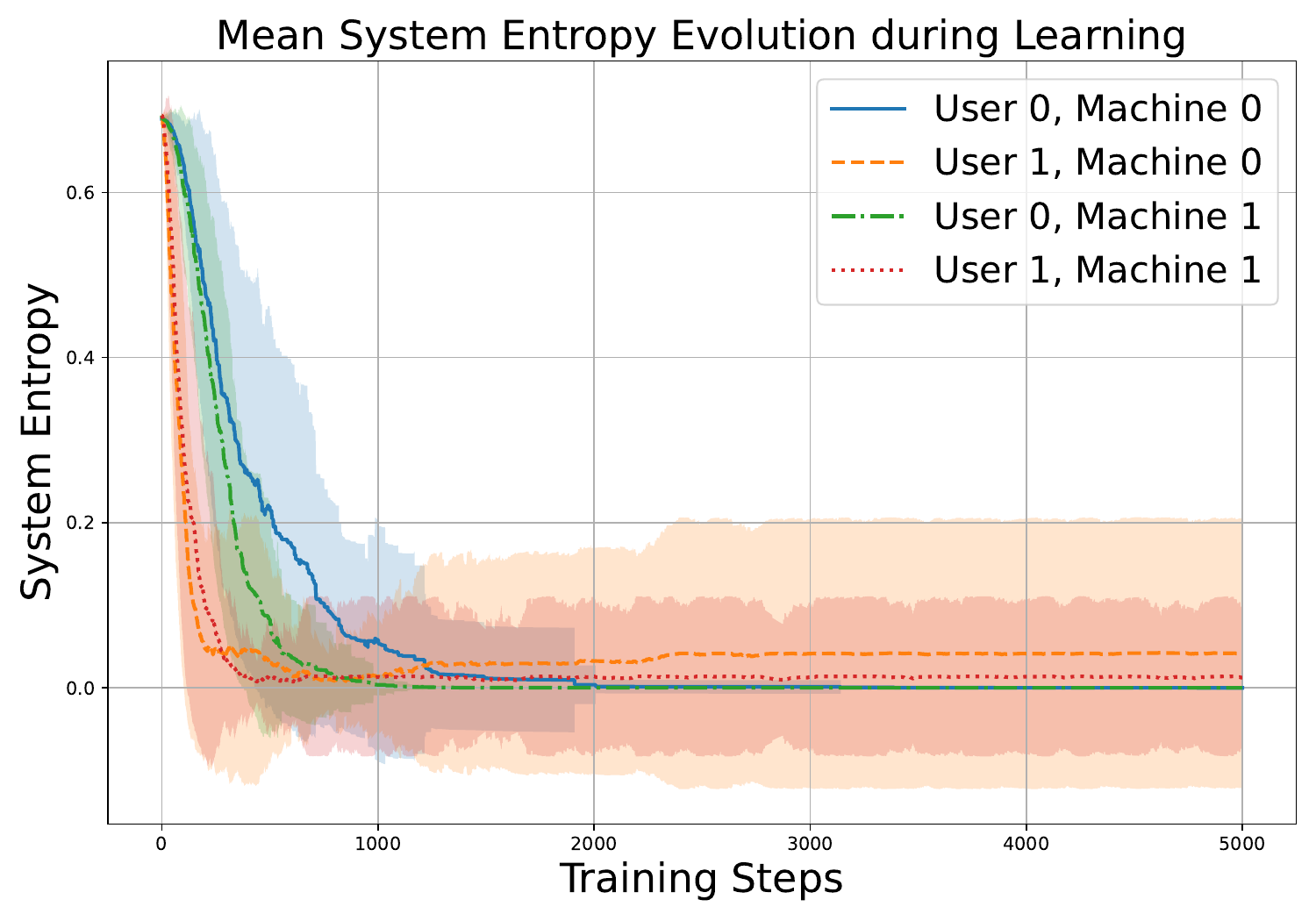}%
}
\vspace{-10pt}
\caption{System entropy of user focused cases. Results are averaged over 50 runs, with the mean shown as the solid line and the standard deviation represented by the shaded area.}
\label{fig: setting1}
\end{figure*}
\section{Evaluation}
To assess whether a user-GenAI pair can function as a single integrated system in the environment, we adopt an information-theoretic approach centered on mutual information and system entropy. System entropy reflects the amount of information in the system, and mutual information quantifies the amount of information shared between random variables, revealing how knowledge of one variable reduces uncertainty about another.

By computing pairwise and joint mutual information shared between the user and the machine over multiple rounds of interaction, we can evaluate the degree of coupling and the temporal coherence in the system.
\vspace{-2mm}
\subsection{Evaluation Setup}

In our simulation, agents learn optimal action policies using Q-learning~\citep{barto2021reinforcement} with a fixed-temperature ($\tau =1$) softmax action selection. Reinforcement learning maintains estimates Q-value of the expected discounted return for each state–action pair and updates them iteratively via Bellman equation. Q-learning does not require knowledge of the environment’s transition probabilities, relying solely on sampled experience. In our implementation, these state dynamics are explicitly parameterised using logistic functions of the current state, action, and contextual variables, enabling flexible control of coupling strengths and noise for experimentation. These functions define stochastic transitions in the simulator but are never used directly by the learning algorithm. To ensure the robustness of the results, the experiment is repeated multiple times under the same settings, and the mean and standard deviation of the outcomes are reported.

\subsection{User-focused scenario}
We first focus on a simplified version of the proposed model where the user's future state $s'$ depends on a combination of the current state, the action $a$ (proportional to current user state), the observation $o$ (proportional to machine state $m$), and the environment $e$. The machine states and environment states are randomly selected during the interaction. We define the user state $\mathcal{S}=\{0,1\}$, machine state $\mathcal{M}=\{0,1\}$, and environment $\mathcal{E}=\{0,1\}$. Here, $s=1$ indicates that the user has good knowledge of a particular task, while $s=0$ represents limited or no task knowledge. Similarly, $m=1$ denotes that the machine is equipped with a stronger ore more effective foundation model, whereas $m=0$ corresponds to a weaker or less capable model. Finally, $e=1$ represents a more complex environment, and $e=0$ denotes a simpler or less informative environment. The current reward depends on the next expected state, i.e.\ when $s=1$, the reward is $\alpha = 1$. The transition probability of the user when the future state $s'$ is 1 is defined in a logistic function:
\begin{equation}
    p(s'|s, o, e)= \sigma[\omega_0 + \omega_s s^t + o^t(\omega_a r^t+\omega_m m^t) + \omega_e e^t + \gamma_s s^t o^t]
    \label{eq: p_x}
\end{equation}
where $\sigma(x)= \frac{1}{1+e^{-x}}$, $\omega_0$ is a bias term controlling the baseline probability, $\omega_s$ is the weight for the current human state, $\omega_a$ and $\omega_m$ are the weights for the request and machine state associated with the observation, $\omega_e$ is the weight for the environmental factor, and $\gamma_s$ is the interaction weight capturing the joint influence of the current human state and observation. Positive weights in Eq.~(\ref{eq: p_x}) can increase the likelihood of staying in state 1, and vice versa. Please refer to the evaluation setting in \ref{tab1}. We choose weights to ensure that action 1 generally promotes reaching $s=1$ (the “goal” state of the user with a good knowledge), while action 0 is less effective. To evaluate the interaction of the user and GenAI, we measure system entropy at each step for each $(s, m)$ pairs as
\begin{equation}
     H(\mathcal{S}, \mathcal{M}) = -\sum_{s, m} P(s, m)\log P(s, m)
     \vspace{-2mm}
\end{equation}
\begin{table}[h]
\small
\center{
\begin{tabular}{cccccc}\hline
Weights & Simple& HC  & LC & HC(Env) & HC(M)\\ \hline\hline
$w_s$ & 2.0 & 1.5 & 0.2 & 1.5& 1.5\\ \hline
$w_a$ & 2.0 & 1.0 & 0.2 & 1.0& 1.0\\ \hline
$w_m$ & -1 & 6.0 & 0.1 & 6.0& 6.0\\ \hline
$w_e$ & 1.0 & 1.5 & 0.1 & 1.5& 1.5\\ \hline
$\tau_m$ & - & 0.0 & 0.0 & 2.0& 2.0\\ \hline
$\tau_a$ & - & 2.0 & 2.0 & 0.0& 5.0\\ \hline
$e_a$ & - & 1.0 & 1.0& 4.0 & 4.0 \\ \hline
$e_s$ & - & 1.0 & 1.0 & 4.0 & 4.0\\ \hline
$e_e$ & - & 1.0 & 1.0 & 2.0 & 2.0\\ \hline
\end{tabular}
}
\vspace{-8pt}
\caption[]{Settings of evaluation. We examine five main configurations: \\
 -- Simple: the user-focused scenario \\
 -- High Coupling (HC): the machine has a strong influence on the user \\
 -- Low Coupling (LC): there is a weak dependency among the user, machine, and environment \\
 -- HC with impactful environment (HC ENV): a more dynamic environment is introduced \\
 -- HC with stable machine (HC M): the machine strongly influences the user, but changes more slowly.}
\label{tab1}
\end{table}

We first demonstrate the system entropy during learning with the user-focused settings, where the user begins interacting with the machine without a prior bias toward any particular action to start in Fig. \ref{fig: u_s_h}. We use system entropy to indicate whether a question can be answered by the machine. Initially, the user—being in a lower knowledge state—chooses strategies to generate prompts with equal probabilities, resulting in high uncertainty. Over time, as the user favours states associated with higher knowledge, strategy 1 yields a higher reward, and the probability of selecting strategy 1 increases, leading to a decrease in system entropy. The high deviation in system entropy across independent runs reflects the stochastic nature of the agent–environment interaction. Variability arises from randomised action selection, environmental state changes, and probabilistic transitions, which amplify during early learning. 

When the user is in a high-knowledge state ($s=1$), their prompts $r$ tend to be more informative and exploratory. Initially, system entropy is low due to prior structure (i.e.\ the user has a clue), but as the machine responds with increasingly rich outputs, the entropy grows. \textit{This demonstrates that the user passes more information through the prompt into the interaction, which might unlock more possibilities, such as questions, subtopics, and nuance.}

In Fig. \ref{fig: u_s_h_skewed}, we set skewed initial values $v(s, m, 0) = 5$ and $v(s, m, 1) = 0$ when $a=0$ and $a=1$, respectively. This makes the user start with an incorrect belief about how machine's state affects the outcome of their actions. The higher variance in system entropy with biased initial knowledge arises because early decisions are skewed toward certain states, amplifying small stochastic differences between runs. Incorrect or incomplete biases must be unlearned at varying rates, and the strong coupling between user and machine states propagates these differences. Reduced exploration further limits the averaging out of environmental randomness, reinforcing early divergences and increasing variability compared to the unbiased case.

For the machine with a low efficient foundation model ($m=0$), learning is fast, so entropy stays low---behaviour becomes confidently deterministic. In contrast, for a machine with a high efficient foundation model ($m=1$), the user must unlearn a false bias, and during this transition, the action probabilities become nearly equal. This results in a temporary spike in entropy, reflecting the user's uncertainty and renewed exploration. \textit{A good understanding of the machine's state is crucial for effective interaction. When the machine behaves efficiently and provides informative responses, it enables the user to receive more useful feedback. This increased information flow not only guides the user's learning but also motivates the user to refine their prompts.} Early user satisfaction or initial system biases can suppress exploration and limit learning; GenAI should encourage curiosity and diverse prompting.

We further illustrate the system entropy dynamics with higher rewards $\alpha=2$ where the user prefers a high knowledge state ($s=1$) more in Fig. \ref{fig: u_s_h_reward}. The user quickly chooses the best strategy and stops exploration, and the system entropy quickly converges. \textit{In GenAI interaction, early satisfaction can suppress exploration. A user who finds initial outputs ``good enough'' may stop prompting or iterating, limiting both the system's expressiveness and the user’s learning. Encouraging iterative prompting helps maintain higher entropy and fosters deeper information exchange.} 
\begin{figure}[h]
    \centering
    \includegraphics[width=1\linewidth]{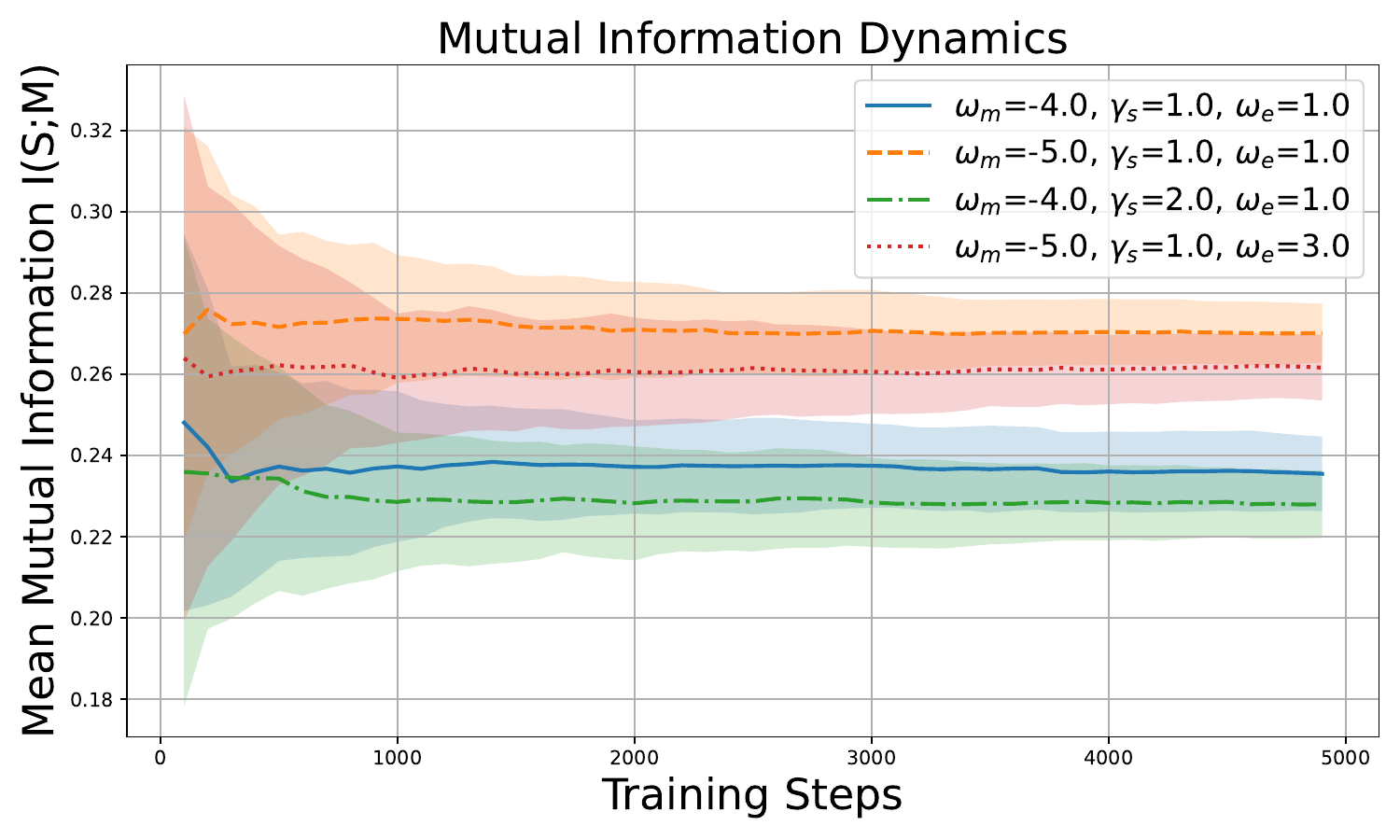}\vspace{-4mm}
    \caption{Mutual information of the user and the machine with unbiased knowledge ($\alpha=1$), plotted against varying weights applied to the user-state transition probabilities. Results are averaged over 50 runs, with the mean shown as the solid line and the standard deviation represented by the shaded area.}
    \label{fig: u_m_mi}
\end{figure}

By tuning $\omega_m$ and $\gamma_s$, it amplifies the machine's role in the user state transition dynamics (more negative $\omega_m$ makes human depend more strongly on machine) and stronger interaction term (larger $\gamma_s$ strengthens the human-machine synergy). In this case, we evaluate the mutual information (MI) between the user and the machine, defined as 
\begin{equation}
    I(S;M) = \sum_{s\in \mathcal{S}}\sum_{m\in \mathcal{M}}p(s,m)\log \frac{p(s,m)}{p(s)p(m)}
    \label{eq: i_s_m}
\end{equation}
where $p(s, m)$ is the joint distribution of user and machine states, and $p(s)$, $p(m)$ are their individual probabilities.
Fig.~\ref{fig: u_m_mi} demonstrates that when the user depends more on the machine and the environment has a stronger influence ($\omega_e$) on the user, MI increases over interaction. This is because increased environmental influence effectively injects noise into the state transitions---obscuring the impact of machine.
It is worth mentioning that when $\omega_m=-4$, $\gamma_s=2$, MI is comparable to the case with $\omega_m=-4$, $\gamma_s=1$. This is because changing $\gamma_s$ only affects the term $s\cdot o$ in Eq. (\ref{eq: p_x}) (where $o$ is proportional to $m$) when $s=1$ and $m=1$, and the sigmoid's diminishing returns can blunt that change.
The variance across runs decreases over time as the user’s policy converges, gradually filtering out environmental noise and stabilizing the dependency on the machine.
\textit{The key takeaway of this case is that when the user and machine interact in a noisy environment---where excessive environmental input initially masks their dependency---the user's learning process can still adapt to this noise and ultimately recover a high level of dependency on the machine.} 
\vspace{-4mm}
\subsection{User--AI--Environment scenario}
We evaluate the interaction between the user, the machine, and the environment. System entropy of this scenario is defined by using joint entropy to quantify the total information in the system:
\begin{equation}
    H(\mathcal{S}, \mathcal{E}, \mathcal{M}) = -\sum_{s, e, m} P(s, e, m)\log P(s, e, m)
\end{equation}
The transition probability of each state follows the conditional distribution below
\begin{equation}
    P(s^{t+1}, e^{t+1}, m^{t+1}|s^{t}, e^{t}, m^{t}) = P_s P_e P_m
\end{equation}
where $P_s$, $P_e$, and $P_M$ are defined as 
\begin{align}
	&P_s (s^{t+1}|s^t, o^t, e^t) = \sigma(\omega_0 + \omega_s s^t + \omega_a o^t + \omega_e e^t + \gamma_s s^t o^t) \\
	&P_e(e^{t+1}|s^t, e^t) = \sigma(\epsilon_0 + \epsilon_s s^t + \epsilon_e e^t +  \gamma_e s^t e^t) \\
	&P_m(m^{t+1}| m^t, r^t) = \sigma(\tau_0 + \tau_m m^t + \tau_r r^t +  \gamma_m m^t r^t)
\end{align}
where $\omega_{0, s, a, e, s}$, $\epsilon_{0,s,e,}$, $\tau_{0,m,r}$, and $\gamma_{s,e,m}$ are weights of the logistic functions. We then define the mutual information of the user and the machine as follows
\begin{equation}
    I(X;Y|Z) = \sum P(x, y, z) \log \frac{P(x, y|z)}{P(x|z)P(y|z)}
\end{equation}
The conditional mutual information of the user and the machine is 
\begin{align}
\small
    I(&s^{t+1};m^{t+1}|s^t, e^t, m^t) \nonumber\\
    &= \sum P(s^t, e^t, m^t) \sum P(s^{t+1}, m^{t+1}|s^t, e^t, m^t)\nonumber\\
    \quad  &\log\frac{P(s^{t+1}, m^{t+1}|s^t, e^t, m^t)}{P(s^{t+1}|s^t, e^t, m^t) P(m^{t+1}|s^t, e^t, m^t)} 
\end{align}
The current reward function is defined as $ u = \alpha s r^2 + m$, where $\alpha$ is a positive weight.

We examine four main configurations: High Coupling (HC), where the machine has a strong influence on the user; Low Coupling (LC), where there is a weak dependency among the user, machine, and the environment; HC with impactful environment (HC ENV), where a more dynamic environment is introduced; HC with stable machine (HC M), where the machine has a strong influence on the user, but the machine changes more slowly (see Table \ref{tab1}). 

\begin{figure}[!htb]
\centering
\subfloat[\footnotesize System entropy]{
\label{fig: u_m_e_h}
\includegraphics[width=0.9\linewidth]{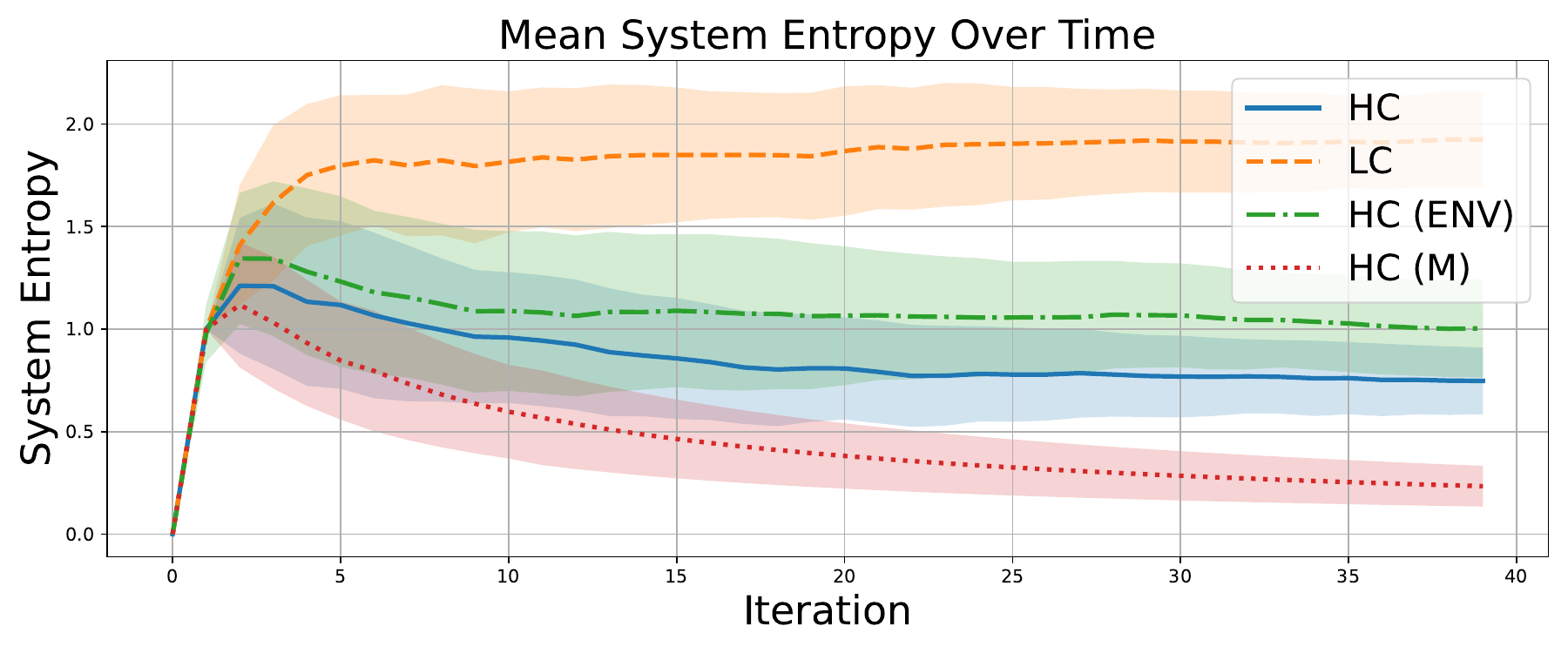}%
}
\hfil
\vspace{-3mm}
\subfloat[\footnotesize Mutual information]{\includegraphics[width=0.9\linewidth]{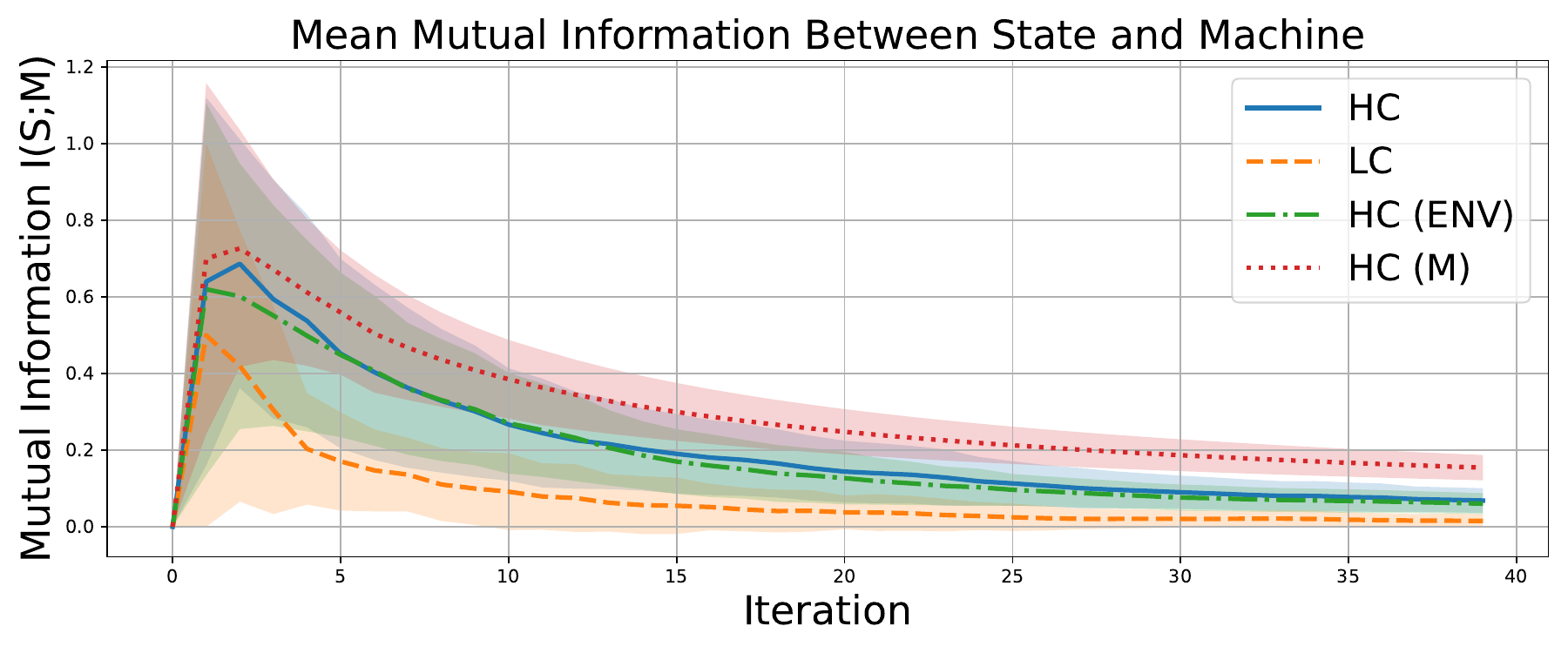}%
\label{fig: u_m_e_mi}}
\vspace{-2mm}
\caption{User-Machine-Environment scenario. Results are averaged over 50 runs, with the mean shown as the solid line and the standard deviation represented by the shaded area.}
\label{fig: setting2}
\end{figure}

We demonstrate the results of the User-Machine-Environment scenario. For the system entropy in Fig. \ref{fig: u_m_e_h}, the initial spike in system entropy from 0 to a high level due to the fixed initial state of the user, machine, and environment. As the agent starts taking actions and the environment becomes involved, the states begin to diversify. Initially, all scenarios lead to higher entropy, because the transitions are more random, which generates more information overall. \textit{A stable machine with low sensitivity to prompts reduces the diversity of possible system trajectories. This leads to lower entropy in the joint system state and restricts the machine's capacity to evolve through user interaction.}
Within one prompt-answer round, the mutual information decreases during the interaction as shown in Fig. \ref{fig: u_m_e_mi}. 
In the HC (ENV) scenario, strong environmental influence increases system entropy by introducing variability in the state transitions. This added uncertainty encourages the user to engage in more exploratory behaviour, as the outcomes are less predictable and more information can potentially be uncovered through interaction.
In contrast, the HC (M) scenario shows a decrease in system entropy, while the mutual information between machine and state remains high. \textit{This suggests that the machine has a strong, stable influence on the system dynamics, making its behaviour more predictable and tightly coupled with the user's. The high mutual information implies that the user and machine become increasingly aligned---operating as a cohesive unit with shared dynamics.} Another interpretation of this result is that an AI system that never changes may hinder co-evolution with the user, reducing potential for creative or adaptive outcomes.


\begin{figure}
    \centering
    \includegraphics[width=0.7\linewidth]{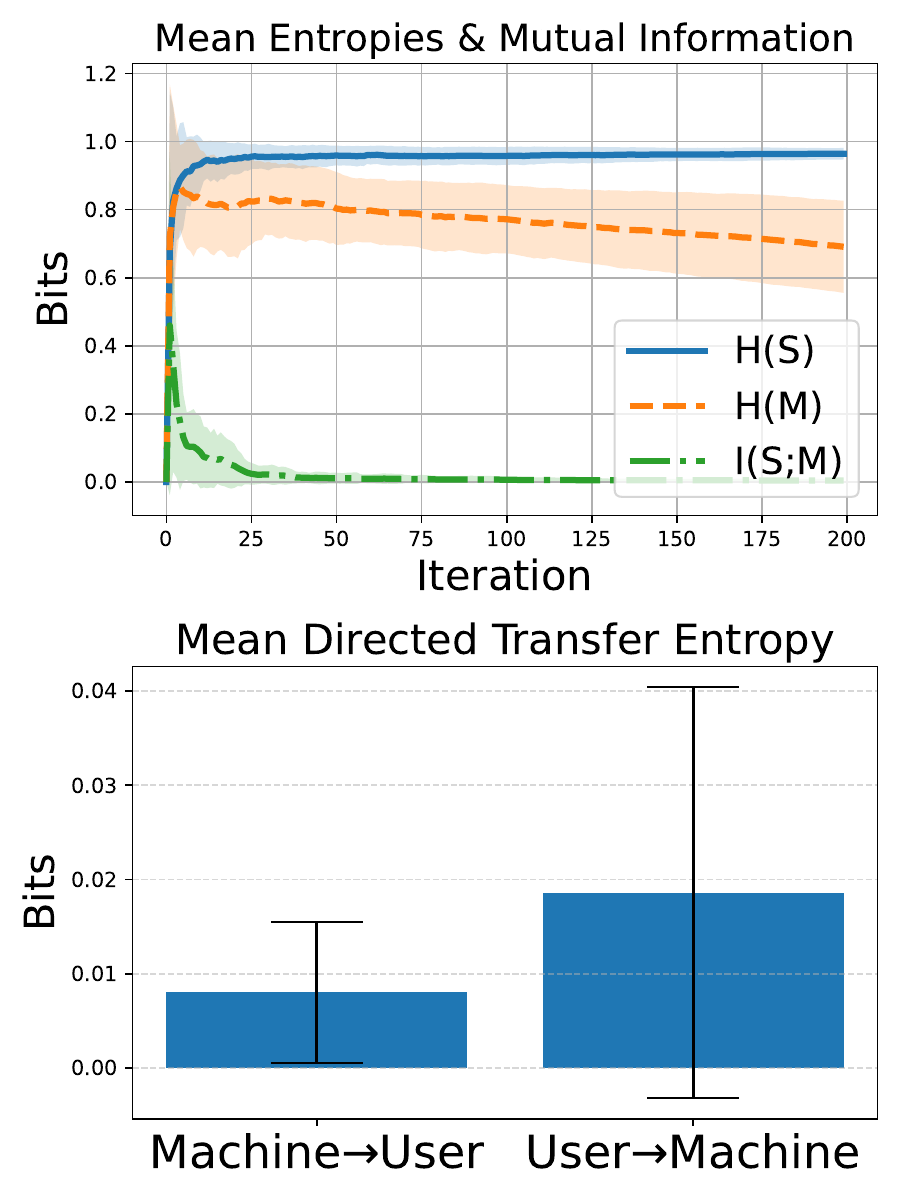}\vspace{-4mm}
    \caption{Machine parasite: entropies and mutual information of user and machine (left) and transfer entropy between user and machine (right). Results are averaged over 50 runs, with the mean shown as the solid line and the standard deviation represented by the shaded area.}
    \label{fig: m_p}
\end{figure}

We set $\omega_m = 0.1$, resulting in the user being barely influenced by the machine. The parameter $\omega_s = -2$ makes the user's state transitions more driven by its own actions, while $\tau_m =5$ causes the machine to be more self-influencing. In Fig. \ref{fig: m_p}, $H(S)$ remains high and $H(M)$ decreases, indicating that the user retains a high level of information while the machine's information quantity shrinks over time. \textit{The machine is using information it extracts from the user to minimise its entropy.} MI peaks during the initial interaction. This occurs because the machine and user are listening to each other and co-varying. The later drop in MI is because once the machine discovers the optimal strategy, it stops tracking the user's ongoing fluctuations. \textit{The machine only briefly pays attention to the user---just long enough to learn how to predict them---then it settles into its own routine, leaving the user's behaviour largely uncoupled.} We also observe that the transfer entropy shows that the machine is heavily influenced by the user, while the user barely learns anything from the machine. This asymmetric flow demonstrates that there is parasitic behaviour in the human--machine interaction.

\vspace{-3mm}
\section{Discussion}

In this work, we propose a flexible approach to model Human--Machine interaction that explicitly accounts for perception, incentives, environmental context, and coupling dynamics. Our results underscore several critical insights. 

First, the initial perception of the machine by the users plays a significant role in shaping subsequent interactions. This finding highlights the importance of calibrating users' general understanding of machines' abilities to guide the system toward desirable equilibria. Second, the reward structure is pivotal. Generous rewards may induce early satisfaction and strong reliance, but suppress exploration. Third, we confirm that environmental factors (as information input) cannot be ignored. Lastly, we demonstrate that a dynamic environment and a well-equipped self-influence machine, powered by an improved foundation model, are important factors for the user and the machine to merge into a single individual. The proposed framework of human--machine interaction demonstrates that close collaboration can yield both benefits (e.g.\ knowledge expansion) and risks (e.g.\ parasitic relationship). Additionally, this model suggests opportunities for designing symbiotic protocols that intentionally loosen coupling to encourage aligned discovery.

While symbiosis between artificial and natural systems has been considered before~\citep{fairclough_closedloopperspective,wang_onthephilosophical}, this has typically been considered only in a positive light with a mutually-beneficial relationship between human and machine. The scenario posited at the outset of this paper (and subsequently modelled; Fig.~\ref{fig: setting2}) shows that this relationship could easily be a parasitic one, with an LLM feeding off the information available to it and the human losing the ability to work effectively without the machine. The proposed model will help to structure Human--AI studies in a rigorous manner by adopting the model we propose, helping to draw out meaningful insights through quantitative measures (i.e.\ via information theory).


The flexibility of our proposed model allows us to study a wide range of different scenarios and extract insights into the behaviour through information-theoretic measures. The main drawback to our approach is that it is not yet calibrated with real-world data. We are planning user studies to calibrate the model, focussed on the request--response cycle ($r^t$ and $o^t(m^t, r^t)$), which would allow us to define the transition probability of the user, $P_s(..)$. This calibrated model will allow us to extract meaningful, real-world, actionable insights, and to predict the impact of various scenarios on humans interacting with an AI system.

For future work developing the model, we plan to extend the model to large-scale human-machine interaction, allowing us to investigate how different forms of symbiosis and individuality emerge in complex ecosystems. Next, we aim to explore the reward function to promote truly cooperative dynamics in human--machine symbiosis, rather than parasitic ones.  
\vspace{-3mm}
\section{Acknowledgement}
Jiejun Hu-Bolz is supported within the project TRACE-V2X, which has received funding from the European Union's HORIZON-MSCA-2022-SE-01-01 under grant agreement No 101131204.

\vspace{-3mm}
\bibliographystyle{apalike}
\bibliography{example} 

\end{document}